\documentclass[fleqn,10pt]{wlscirep}
\usepackage[utf8]{inputenc}
\usepackage[T1]{fontenc}
\usepackage{bm}
\usepackage{siunitx}
\usepackage{soul}

\usepackage{amsmath,hyperref}
\usepackage{amsfonts,amssymb}
\usepackage{lineno}
\usepackage{ulem}
\hypersetup{
     colorlinks=true,
     linkcolor=blue,
     filecolor=blue,
     citecolor = blue,      
     urlcolor=blue,
     }
\usepackage{tcolorbox}
\usepackage{tabularx}
\usepackage{array}
\usepackage{colortbl,ulem}
\tcbuselibrary{skins}

\newcommand\redout{\bgroup\markoverwith
{\textcolor{red}{\rule[.5ex]{2pt}{0.4pt}}}\ULon}

\newcolumntype{Y}{>{\raggedleft\arraybackslash}X}

\tcbset{tab1/.style={fonttitle=\bfseries\large,fontupper=\normalsize\sffamily,
colback=yellow!10!white,colframe=red!75!black,colbacktitle=yellow!40!white,
coltitle=black,center title,freelance,frame code={
\foreach \n in {north east,north west,south east,south west}
{\path [fill=red!75!black] (interior.\n) circle (3mm); };},}}

\tcbset{tab2/.style={enhanced,fonttitle=\bfseries,fontupper=\normalsize\sffamily,
colback=yellow!10!white,colframe=red!50!black,colbacktitle=yellow!40!white,
coltitle=black,center title}}

\title{Hedgehog topological defects in 3D amorphous solids}
\author[1,*]{Arabinda Bera}
\author[1$\dagger$]{Alessio Zaccone}
\author[2$\ddagger$]{Matteo Baggioli}
\affil[1]{Department of Physics ``A. Pontremoli", University of Milan, via Celoria 16, 20133 Milan, Italy}
\affil[2]{Wilczek Quantum Center, School of Physics and Astronomy, Shanghai Jiao Tong University, Shanghai 200240, China \& Shanghai Research Center for Quantum Sciences, Shanghai 201315, China}
\affil[$\ddagger$]{ 
\color{blue}arabinda.bera@unimi.it\color{black}; \color{blue}alessio.zaccone@unimi.it\color{black};
\color{blue}b.matteo@sjtu.edu.cn\color{black}\vspace{0.2cm}}

\begin{abstract}
\textbf{The underlying structural disorder renders the concept of topological defects in amorphous solids difficult to apply and hinders a first-principle identification of the microscopic carriers of plasticity and of the regions more prone to structural rearrangements (``soft spots''). Recently, it has been proposed that well-defined topological defects can still be identified in glasses, and correlated to local and global plasticity, by looking at the eigenvector field or the particle displacement field. Nevertheless, all the existing proposals and analyses are only valid in two spatial dimensions. In this work, we propose the idea of using hedgehog topological defects to characterize the plasticity of 3D glasses and to geometrically predict the location of their soft spots. We corroborate our proposal by simulating a Kremer-Grest 3D polymer glass, and by using both the normal mode eigenvector field and the displacement field around large plastic events. \color{black}Contrary to the 2D case, the sign of the topological charge defined from the eigenvector field is ambiguous and the geometry of the topological defects, whether radial or hyperbolic, plays a fundamental role in 3D. In fact, we find that the topological hedgehog defects relevant for plasticity are those exhibiting hyperbolic geometry, resembling the saddle-point structure of 2D topological defects with negative winding number (anti-vortices). \color{black} Our results confirm that a topological characterization of plasticity in glasses is feasible, and provide a concrete realization of this program in 3D amorphous systems.}
\end{abstract}

\begin{document}
\flushbottom
\maketitle
\thispagestyle{empty}

 Topological defects, such as dislocations and disclinations, play a pivotal role for the description of mechanical failure in crystalline solids since they provide the microscopic origin of their plasticity \cite{Taylor,Polanyi,Orowan}. Topological defects also provide the building blocks for the theory of crystal melting in two dimensions  \cite{kosterlitz1973,Halperin1978,Young1979}. Despite several attempts \cite{SPAEPEN1978207,PhysRevLett.42.1541,doi:10.1080/01418617908234879,10.1063/1.326364,POPESCU1984317,EGAMI1984499,SHI199368,PhysRevLett.43.1517,doi:10.1080/01418618008243894,ACHARYA20171,PhysRevB.28.5515,doi:10.1080/01418618108235816}, topological defects associated to translational symmetry (\textit{e.g.}, dislocations) cannot be formally defined within the structure of amorphous solids such as glasses, because of the absence of translational long-range order (or quasi-long range order in 2D systems). This difficulty remains the main cause for the lack of a microscopic theory of plasticity and mechanical deformations in glasses \cite{hufnagel2016deformation,rodney2011modeling,zaccone2023theory}. 

More in general, the definition and identification of ``defects'' constitute the main obstacle for establishing a robust causal relation between structure and dynamics in amorphous solids (see \textit{e.g.} \cite{wang2019dynamic,cheng2011atomic}). So far, most of the approaches to solve this problem are still based on ad-hoc coarse-grained descriptions \cite{RevModPhys.90.045006}, vibrational soft modes \cite{PhysRevE.89.042304,manning2011vibrational,ding2014soft,chen2011measurement,kapteijns2020nonlinear,tanguy2010vibrational,smessaert2014structural}, shear transformation zones revisited under the lens of anharmonic quasi-localized modes  \cite{zylberg2017local,richard2021simple,gartner2016nonlinear,ruan2022predicting,xu2018predicting,PhysRevMaterials.5.025603}, phenomenological structural indicators \cite{richard2020predicting,widmer2008irreversible} or machine-learning algorithms \cite{Ronhovde2011,Ronhovde2012,fan2021predicting,ciarella2023finding}. 

Two recent works \cite{baggioli2021,wu2023} (see also \cite{Baggioli2023}) have shown that not only topological concepts can be applied (albeit with proper generalizations) to amorphous solids, as done for vortices in superfluids or dislocations in crystals, but that they can be also extremely useful to predict their mechanical properties both on a local and a global scale. 

A first approach \cite{baggioli2022} is based on the idea of continuous Burgers vector \cite{RevModPhys.80.61} applied to the displacement field, and has been shown to be able to accurately locate plastic events under deformation and the yielding point \cite{baggioli2021}. It is likely that this concept is related to the geometric charges previously discussed as the carriers of plastic screening in glasses \cite{PhysRevE.104.024904}. The second approach \cite{wu2023} relies on the identification of vortex-like defects in the eigenvector field that strongly correlate with the plastic spots and display tight connections with the widely used concept of shear transformation zones \cite{desmarchelier2024topological}. This second type of vortex-like topological defects have been recently experimentally identified in a 2D colloidal glass \cite{vaibhav2024experimental}, confirming the simulation results.

Importantly, all the proposals and analyses so far apply only to two-dimensional (2D) amorphous solids and lack direct application to more realistic 3D scenarios. In \cite{Cao2018}, disclination-like defects, defined using polytetrahedral order \cite{nelson1989polytetrahedral}, were considered in a 3D experimental granular matter system and were shown to correlate with plasticity. In \cite{bera2024soft}, a 2D slicing method has been used to study vortex-like topological defects in a 3D simulation polymer glass revealing a strong clustering tendency of negative charged topological defects right before large plastic events. 

As of today, a full-fledged definition of topological defects in 3D glasses and a direct proof of their connection to plasticity in three dimensions are still lacking. In this work, we present a solution to these problems. 

\section*{Hedgehog Topological Defects and 3D polymer glass model}
 Given the large variety of topological defects that can be defined within a three-dimensional vector field (point defects, line defects and surface defects) \cite{kleinert1989gauge,nelson2002defects}, we suggest that hedgehog topological defects (HTD) are the simplest choice to start with. We adapt the definition of HTDs from nematic liquid crystals \cite{Kleman,kleman12006,holbl2023} and Heisenberg magnets \cite{fumeron2023introduction,lau1989,holm1994,berg1981} to the eigenvector and displacement fields of a 3D simulation glass. 

For a 3D unit vector field $\vec{s}(x)$ we define the hedgehog topological charge as \cite{Kleman,kleman12006,holbl2023,senyuk2013}
\begin{equation}\label{eq2}
Q = \frac{1}{4\pi} \int_{\Sigma} d^2x~\vec{s} \cdot (\partial_1 \vec{s} \times \partial_2 \vec{s}),
\end{equation}
where $\Sigma$ is an arbitrary closed 2D surface parametrized by coordinates $(x_1, x_2)$. $Q$ represents a topological invariant in the sense that it does not change under continuous and smooth deformations of the manifold. \color{black}
We notice that, whenever the system is invariant under the parity transformation $\vec{s}\rightarrow - \vec{s}$ (\textit{e.g.}, nematic liquid crystals), then the sign of the topological charge $Q$ is ambiguous and hence not physical.\color{black}

\begin{figure*}[t]
    \centering
    \includegraphics[width=16cm]{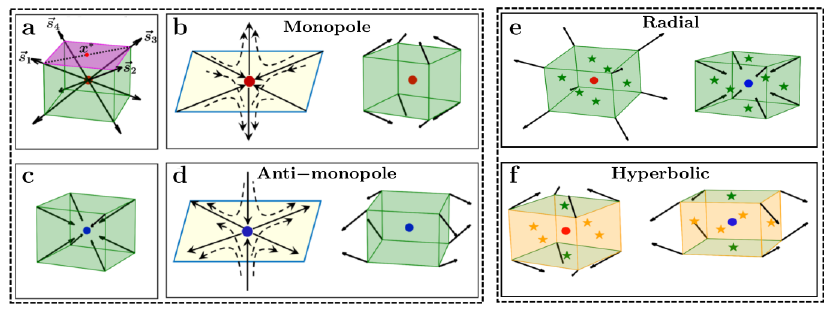}
    \caption{{\bf Schematic representation of 3D point defects.} {\bf a}~Discrete vector at a simple cubic lattice corners illustrating a radially outward field for a monopole with $Q=+1$ (\color{red}\textbf{red}\color{black}). A vertex $x^*$ is shown where $\vec{s}_1$, $\vec{s}_2$, $\vec{s}_3$ and $\vec{s}_4$ field at four corners of the vertex. {\bf b}~Continuous vector field (left) and discrete vector field (right) for a monopole. {\bf c}~Radially inward field representing an anti-monopole with $Q=-1$ (\color{blue}\textbf{blue}\color{black}). {\bf d}~Continuous (left) and discrete (right) vector fields for $Q=-1$. \textcolor{black}{{\bf e}~Radial defects with $Q=\pm 1$ are shown along with the 2D winding numbers of the projected vectors onto the faces of a cube. Negative 2D winding number ($q_{2D}=-1$) is indicated in orange, positive ($q_{2D}=+1$) in green. Stars mark the locations of the 2D point defects, while the corresponding surfaces are colored according to their winding number. {\bf f}~Hyperbolic defects with $Q=\pm 1$ are displayed with projected vectors and their associated winding numbers, using the same color scheme as in panel {\bf e}.}
}
    \label{fig1}
\end{figure*}
We implement the concept of 3D point  defect on a lattice following the methods in refs. \cite{holm1994,berg1981}.
From our simulations, we first obtain the unit vector field ${\vec{s}}_{l,m,n}$ at each lattice grid point $(l,m,n)$ in a simple cubic lattice. More details are provided in the Methods section. Let $\vec{s}_1$, $\vec{s}_2$, $\vec{s}_3$, and $\vec{s}_4$ be the unit vectors of the field associated with one face of the cube, where $\vec{s}_1$, $\vec{s}_2$, and $\vec{s}_3$ are positioned at the corners of triangle $i$ on that face, chosen such that the sequence $1-2-3-1$ corresponds to a counterclockwise rotation along the outward normal to the surface of the triangle. See Figure~\ref{fig1}a for a representation of these vectors on the example of a radially outgoing field (monopole). The topological charge $Q$ enclosed by the unit cube is then expressed in a discrete fashion as \cite{holm1994,berg1981},
\begin{equation}\label{eq1}
Q=\sum_{x^*}q(x^*),
\end{equation}
where the sum is carried over all $6$ vertices of the unit cube and $x^*$ is the vertex associated with the four vectors $\vec{s}_1$, $\vec{s}_2$, $\vec{s}_3$ and $\vec{s}_4$ (as shown in Figure~\ref{fig1}a). Here, $q(x^*)$ is expressed as
\begin{equation}\label{eq1}
q(x^*)=\frac{1}{4\pi}[(\sigma_s A)(\vec{s}_1,\vec{s}_2,\vec{s}_3)+(\sigma_s A)(\vec{s}_1,\vec{s}_3,\vec{s}_4)],
\end{equation}
where $(\sigma_s A)(\vec{s}_1,\vec{s}_2,\vec{s}_3)$ is the signed area of the spherical (non-Euclidean) triangles with corners $\vec{s}_1$, $\vec{s}_2$ and $\vec{s}_3$. 
The sign of the area can be calculated as ${\sigma_s=\rm{sign}}\{\vec{s}_1\cdot[\vec{s}_2\times \vec{s}_3]\}$. Finally, the value of $A$ is calculated as \cite{holm1994,berg1981},
\begin{equation}\label{eq1}
\cos \left (\frac{A}{2}\right )=\frac{1+\vec{s}_1\cdot \vec{s}_2+\vec{s}_2\cdot \vec{s}_3+\vec{s}_3\cdot \vec{s}_1}{\sqrt{2(1+\vec{s}_1\cdot \vec{s}_2)(1+\vec{s}_2\cdot \vec{s}_3)(1+\vec{s}_3\cdot \vec{s}_1)}}.
\end{equation}
A visual representation of this algorithm for the case of hedgehog defects with charges $Q=\pm 1$ (monopole and anti-monopole) is presented in \textcolor{black}{Figure~\ref{fig1}a-d.}

\color{black}To identify the hedgehog defects, we use the cubic lattice construction as described above. We further analyze and classify the hedgehog defects based on the geometric structure of the vector field around their core \cite{senyuk2013,lavrentovich1986}.  The vector fields for an ``ideal'' radial (R) defect (Figure~\ref{fig1}a) and an ``ideal'' hyperbolic (H) defect (Figure~\ref{fig1}b) are given by \cite{lavrentovich1986}:
\begin{align}
 \vec{s}_R(x,y,z) = \{x,y,z\}(x^2+y^2+z^2)^{-1/2},\qquad \vec{s}_H(x,y,z) = \{-x,-y,z\}(x^2+y^2+z^2)^{-1/2},\label{eee}
\end{align}
where both defects are located at the origin $(0,0,0)$. Importantly, both configurations in Eq. ~\eqref{eee} are associated to a topological charge $Q=+1$, proving that the sign of $Q$ does not fully characterize the defect properties in 3D since it is insensitive to the geometric structure. This is different to the case in 2D, where the sign of the winding number characterizes also the geometry of the defect: radial for vortices with positive charge, while hyperbolic for anti-vortices with negative charge. We notice that inverting the field direction in Eq. \eqref{eee} ($\vec{s} \rightarrow - \vec{s}$) results in switching the sign of the topological charge $Q$ (as shown in Figure~\ref{fig1}c,d). Nevertheless, this operation does not alter the radial or hyperbolic nature of the defect.

To practically characterize the geometric structure of the hedgehog defects, we project the 3D vector field onto the six faces of a cubic cell and then we compute the 2D winding number on each face as $q_{2D}=(1/2\pi)\oint \vec{\nabla} \theta \cdot \vec{d\ell}$, where $\theta$ is the phase angle and $\vec{d\ell}$ is the line element along a closed square loop. For an ideal radial hedgehog with $Q=+1$, then $q_{2D}=+1$ on all faces (Figure~\ref{fig1}e). In contrast, an ideal hyperbolic defect with $Q=+1$ exhibits $q_{2D}=+1$ on two faces and $q_{2D}=-1$ on the remaining four (Figure~\ref{fig1}f). This distinction allows us to clearly differentiate between radial and hyperbolic defects. 

In our simulation system, the situation is more complicated. In fact, for an arbitrary defect core, we obtain various combinations of $q_{2D}$, which typically takes three distinct values, $-1$, $0$ and $+1$ (for concrete examples, see Fig. \ref{figSI4} in the SM) and do not necessarily align with the ideal scenarios presented above. 
For each hedgehog defect, we further define the parameter $N_s$ that is the number of faces with $q_{2D}=-1$ (saddle-like). Clearly, for the ideal radial hedgehog $N_s=0$, and for the ideal hyperbolic hedgehog $N_s=4$. Then, we define the defects to be radial if $N_s=0$ and hyperbolic if $N_s>0$, while the value of $N_s$ could be thought as how exact the hyperbolic nature is. This classification is motivated by the inherent association of hyperbolic structures with saddle-like features onto the projected plane.\color{black}

We simulate a 3D polymer glass system using the Kremer-Grest model \cite{Kremer} with $N = 10{,}000$ monomers of alternating masses of $m_1 = 1$ and $m_2 = 3$ \cite{PhysRevB.102.024108} interacting via the Lennard-Jones (LJ) potential ($u_{\text{LJ}}$) and the finite extensible nonlinear elastic (FENE) potential ($u_{\text{FENE}}$). We set the dihedral bending rigidity of the bonds to zero, as appropriate for flexible chains \cite{palyulin2018}. More details can be found in the Method section.

We define the $3N \times 3N$ Hessian matrix, with elements written in terms of the total potential energy $u(r)=u_{\text{LJ}}(r)+u_{\text{FENE}}(r)$,
\begin{equation}\label{eq_hessian}
H_{ij}^{\alpha\beta} = \frac{1}{\sqrt{m_i m_j}} \frac{\partial^2 u(\vec{r})}{\partial {r}_i^{\alpha} \partial {r}_j^{\beta}}.
\end{equation}
Here, $i$ and $j$ are particle indices, while $\alpha$ and $\beta$ denote the spatial directions ($x$, $y$, $z$). Here $m_i$ represents the mass of the $i$-th particle, and $r_{i}^{\alpha}$ is the $\alpha$-component of the position vector of the $i$-th particle.

The eigenvalues $\lambda_k$ (or equivalently the eigenfrequencies $\omega_k = \sqrt{\lambda_k}$) and their associated eigenvectors are obtained by direct diagonalization of the Hessian matrix. In the rest of the manuscript, we will use the symbols $\omega_k$ and $\omega$ interchangeably.

\section*{Soft spots from the topology of the eigenvector field}
We start by applying our algorithm to the 3D eigenvector field of the simulated polymer glass system. To determine the HTDs, we derived the smoothed eigenvector field on a simple cubic lattice (see Methods for details). For direct visualization, we present a snapshot of the smooth eigenvector field that exhibits an hedgehog anti-monopole and a monopole in the Methods, Figure \ref{figSI1}.

By scanning the eigenvector field at different frequencies, we have verified that the total number of HTDs scales at low frequency (up to $\omega \approx 4$) as $N_{\text{HTD}} \propto \omega^2$ (shown in Supplementary Material, Figure \ref{figSI3}b) and correlates with the corresponding density of states, which exhibits Debye scaling, $D(\omega) \propto \omega^2$ (see SM, Figure \ref{figSI3}b) at low frequency. \color{black} We notice that this scaling does not align with a simple argument based on counting stationary points in plane wave eigenvectors presented in \cite{wu2023}. We found that such argument fails in our context since it reproduces only hedgehog defects with radial geometry, a scenario that appears incompatible with our simulation results. It would be interesting to understand this discrepancy better in the near future. Along these lines, we notice that a recent study \cite{wu2024} also reports a quadratic scaling of line defect density in 3D glasses.\color{black}

After confirming the validity of the 3D algorithm presented above, we investigate the physical properties of the HTDs, and in particular whether they bear any connection with plastic soft spots. In order to do so, we follow \cite{C4SM01438C} and we identify the soft spots in the 3D model glass system using the softness parameter
\begin{equation}\label{softdef}
    \phi_i = \frac{1}{N_m} \sum_{j=1}^{N_m} \frac{|{\bf e}^{(i)}_j|^2}{m_i\omega_j^2},
\end{equation} 
where $N_m$ is the number of low frequency modes used in the analysis and ${\bf e}^{(i)}_j$ is the eigenvector of the $j$-th mode, corresponding to the $i$-th particle. To define $N_m$, we have considered modes up to $\omega=5$. Finally, the plastic soft spots are identified using the locations of the $5\%$ of particles having higher values of $\phi_i$.
  
\begin{figure*}[t]
    \centering
\includegraphics[width=16cm]{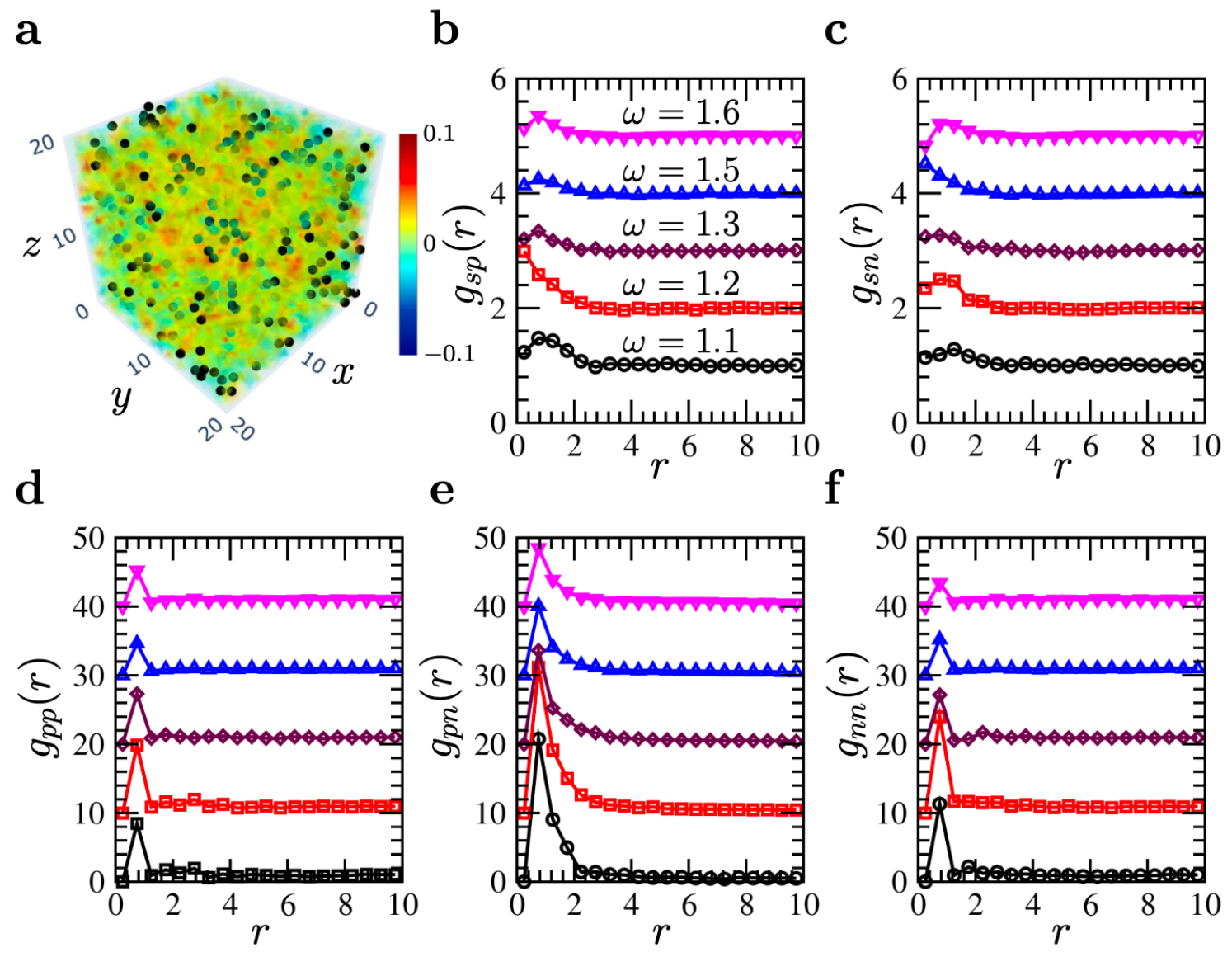}
    \caption{{\bf Correlation between the soft spots and the topological defects.} {\bf a}~The color map visualizes the 3D average topological charge density. The black symbols are the locations of soft spots, identified using the softness parameter, Eq. \eqref{softdef}. {\bf b-f}~Pair correlation functions between HTDs: monopoles (p) and anti-monopoles (n), and plastic soft spots (s) for different frequencies. Note that the curves for $\omega>1.1$ are shifted upward by constant factors for better visualization in this and all other similar data sets. See SM, Fig.~\ref{figSI4}, for extended data.}
    \label{fig2}
\end{figure*}
\begin{figure*}[t]
    \centering
\includegraphics[width=16cm]{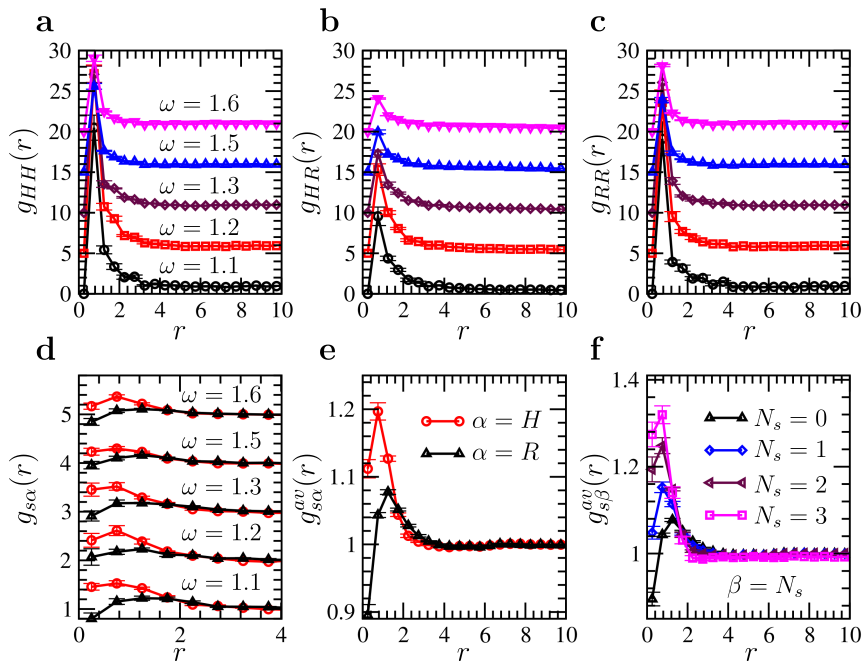}
    \caption{\textcolor{black}{{\bf Correlation between soft spots and topological defects with different geometric structure.} {\bf a-d} Pair correlation functions between HTDs with hyperbolic (H) and radial (R) nature and plastic soft spots (s) for different frequencies. The curves corresponding to $\omega>1.1$ are shifted upward by constant factors for improved visualization in this and all similar datasets. {\bf e} The average correlation function between soft spots and hyperbolic (H) and radial (R) defects. {\bf f} The average correlation function between soft spots and topological defects corresponding to different $N_s$ sectors. Error bars are included in all datasets.}}
    \label{fig3}
\end{figure*}
In Figure~\ref{fig2}a, we show a 3D color map indicating the intensity of the average topological charge density (see SM for details). \color{red}\textbf{Red} \color{black} and \color{blue}\textbf{blue} \color{black} colored regions represent therefore areas with higher concentrations of positive and negative topological charge $Q$, respectively. In the same box, we provide the locations of the plastic soft spots using black symbols. In order to test the correlation between HTDs and plasticity, in Figure~\ref{fig2}b-g, we present various pair correlation functions between HTDs (p), anti-monopole HTDs (n) and plastic soft spots (s) for different frequencies.

In Figure~\ref{fig2}b-c, the correlation between HTDs and soft spots is shown. Our results reveal that both positive and negative defects exhibit strong correlations with soft spots at short and intermediate distances at low frequency, that nevertheless vanish quickly by increasing $\omega$ (see SM, Figure \ref{figSI4}). This outcome presents marked differences with the 2D case \cite{wu2023}, where soft spots predominantly correlate with negative defects at short distances. \color{black}Here in 3D, we find that the sign of the topological charge $Q$ does not play an important role. In fact, this could have been anticipated by looking at the definition of the topological charge $Q$, Eq. \eqref{eq2}, and by recalling that the sign of the eigenvector field in the normal mode analysis is arbitrary. Since inverting the sign of the eigenvector field is not physical, it follows that the sign of the topological charge $Q$, identifying the vector field $\vec{s}$ with the eigenvector field, is also not physical. This explains our findings that, up to statistical fluctuations, do not present any difference between positive and negative charges.\color{black}

Moreover, HTDs with same charge also display a peak in their spatial correlation at approximately the same distance, see Figure~\ref{fig2}d and Figure~\ref{fig2}f, indicating a widespread tendency to clustering of HTDs. This may be explained with the fact that HTDs tend to nucleate more in regions of significant nonaffine motions \cite{baggioli2022}, which are spatially localized \cite{PhysRevE.72.066619}. Hence, it is statistically more likely to have charges, even with same sign, next to each other. Finally, in Figure~\ref{fig2}e, we observe a very strong short-distance correlation between monopole and anti-monopole defects that suggests that defects with opposite charges attract each other and tend to pair together to minimize the local topological charge. This correlation is very strong at low frequency and becomes weaker at higher $\omega$. 

\color{black}As discussed above, in 3D the sign of the charge $Q$ for point defects defined from the eigenvector field is arbitrary. Hence, differently from the 2D case, it is not helpful in identifying the defects responsible for plasticity. In order to solve this problem, we further differentiate the hedgehog topological defects (HTDs)  based on their geometric structure as explained in the previous Section. Interestingly, our simulations do not reveal the presence of only ideal radial or hyperbolic defects (Figure~\ref{fig1}e-f). Instead, we observe a range of intermediate configurations characterized by different values of $N_s$, with representative examples shown in Supplementary Figure~\ref{figSI4}a-f.\color{black}

\textcolor{black}{In Figure~\ref{fig3}a-c, we present the spatial correlation functions for hyperbolic-hyperbolic, hyperbolic-radial, and radial-radial defect pairs. We observe that hyperbolic-hyperbolic and radial-radial correlations are stronger compared to hyperbolic-radial correlations, indicating a natural tendency for defects of the same geometric type to be spatially correlated. In Figure~\ref{fig3}d, we show the correlation functions between soft spots and defects, $g_{s\alpha}(r)$, where $\alpha = H, R$. At low frequencies, the correlation with hyperbolic defects remains consistently stronger than with radial defects, as evident from Figure~\ref{fig3}d. To further quantify this behavior, we compute the frequency-weighted average correlation function \cite{wu2024}:
\begin{equation}\label{correlation}
g_{s\alpha}^{\text{av}}(r) = \frac{\sum_{k} g_{k,s\alpha}(r)/{\omega}_k^2}{\sum_{k} 1/{\omega}_k^2},
\end{equation}
where $\alpha = H, R$, and the sum runs over the selected frequency range. The $\omega^2$ weighting accounts for the quadratic scaling of the defect density at low frequencies. We consider a frequency range $\omega_k \in [1,\omega_c]$ with $\omega_c=3.5$, discarding $\omega_k < 1$ due to large fluctuations and the significantly lower number of hedgehog defects ($N_d \approx 0-50$) in this range.}

\textcolor{black}{The averaged correlation functions $g_{s\alpha}^{\text{av}}(r)$ for $\alpha = H, R$ are shown in Figure~\ref{fig3}e, clearly demonstrating that hyperbolic defects are more strongly correlated with soft spots than radial defects. In Figure~\ref{fig3}f, we present the averaged correlation functions $g_{s\alpha}^{\text{av}}(r)$ for different values of $N_s$ (the number of saddle-like surfaces). The results reveal that the correlation strength increases with $N_s$, indicating that defects with more saddle-like features are more strongly associated with soft spots. We further verify this trend by varying the frequency range used in the averaging procedure. As the frequency range $\omega_c$ increases, the first peak height in $g_{s\alpha}^{\text{av}}(r)$ decreases, as shown in the supplementary figure~\ref{figSI6}a. This suggests that the correlation is strongest at lower frequencies, in agreement with previous reports in the literature \cite{wu2023}. Additionally, we compute the averaged correlation function for defects categorized by their topological charge $Q=+1$ and $Q=-1$ and find no significant differences, as shown in the supplementary figure~\ref{figSI6}b.}

\textcolor{black}{These findings highlight that, for 3D point defects in the eigenvector field, geometry, and not only topology, plays a fundamental role in identifying the carriers of plasticity. In particular, our results suggest that the topological defects relevant to plasticity are those with hyperbolic geometry, aligning with the results in 2D systems \cite{wu2023,desmarchelier2024} where this hyperbolic structure is associated to anti-vortices with $q_{2D}=-1$.}

\section*{Predicting plasticity from the topology of the displacement vector}
Inspired by the proposal of looking at topological defects in the dynamical displacement field $\vec{u}$ \cite{baggioli2021}, rather than in the eigenvector field, we now apply the same 3D algorithm to find HTDs in $\vec{u}$. We notice that there is no direct relation between the displacement field and a single eigenvector $\mathbf{e}_k$, despite $\vec{u}$ can be decomposed onto the complete basis set of eigenvectors. Therefore, a correlation between topological defects in the eigenvector field and in the displacement field is not obvious. Nevertheless, around plastic events, it is known that the dynamical response is dominated by a few localized and low frequency modes (see \textit{e.g.} \cite{yang2020complexity}), providing a possible connection between the topology of the low-frequency eigenvectors and that of the displacement field.

In our study, we apply quasi-static shear to our system using the athermal quasi static (AQS) protocol \cite{lemaitre2006sum} with strain step $\delta \gamma=10^{-3}$. The stress-strain curve is presented in the Methods section (Figure \ref{figSI1}c) and displays a yielding instability at $\gamma^*\approx 0.099$. For presentation purposes, here we focus on four plastic events, including the yielding point, and consider the non-affine displacement field at these specific strain values. In order to test the universality of our findings, we define the soft spots using two different methods: (I) the softness parameter defined in Eq.~\eqref{softdef} (and used in the previous analysis) and (II) the value of the particle level non-affine displacement field $\vec{u}_{\text{NA}}$. Within this second method, we identify the plastic soft spots by looking at the locations of the $5\%$ of particles having higher values of $|\vec{u}_{\text{NA}}|$.

\begin{figure*}[t]
    \centering
\includegraphics[width=16cm]{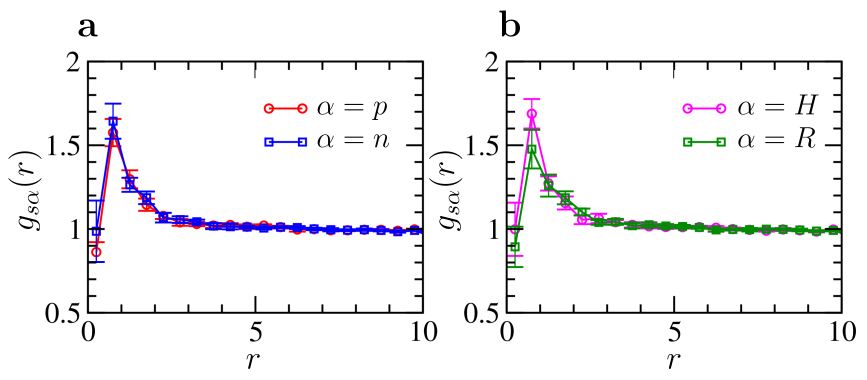}
     \caption{{\bf Correlation between soft spots and HTDs.} {\bf a}~\textcolor{black}{Spatial correlation functions between soft spots (identified using the softness parameter, Eq. \eqref{softdef}) and HTDs within the non-affine displacement field. The correlations for HTDs with topological charge $Q=+1$ ($-1$) are shown in red (blue), respectively. {\bf b}~Spatial correlations between soft spots and HTDs with hyperbolic and radial geometrical nature. Both data sets are averaged over different $\gamma$ values (with error bars), corresponding to plastic events as indicated in Figure~\ref{figSI1}c.}}
    \label{fig4}
\end{figure*}

In order to test the predictability of the plastic regions from pure static configurations, we obtain the soft spots using method (I) applied just before the plastic events. We \textcolor{black}{investigate} the spatial correlation between HTDs within non-affine displacement field and the soft spots identified with the softness parameter $\phi_i$ (method I) \textcolor{black}{across various plastic events. We analyzed $10$ plastic events, for which the stress drops in the stress-strain curve are indicated in the Methods section and shown in Figure~\ref{figSI1}c. In Figure~\ref{fig4}a, we present the average correlations between soft spots (Method I) and HTDs with topological charges $Q=+1$ and $Q=-1$.} From there, it is evident that both positive and negative HTDs within the displacement field display a clear correlation with the soft spots that is rather localized around $r \approx 1$. Once again, no marked distinction between negative and positive defects is observed. \textcolor{black}{We now distinguish the HTDs based on their geometric projections, classifying them as hyperbolic or radial, as discussed earlier. In Figure~\ref{fig4}b, we present the correlations between soft spots and hyperbolic defects ($g_{sH}(r)$), as well as soft spots and radial defects ($g_{sR}(r)$). We observe that the correlation for hyperbolic defects is stronger than for radial defects at shorter distances. These findings further support the observations discussed in the previous sections.} 

\begin{figure*}[t]
    \centering
    \includegraphics[width=\linewidth]{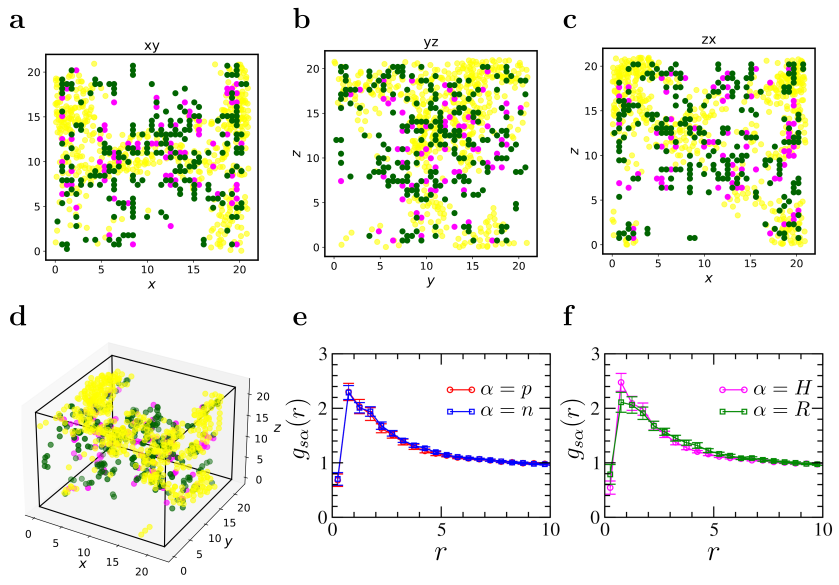}
    \caption{{\bf Visualization and correlation between HTDs and non-affine regions from displacement field.} {\bf a-c}~The location of HTDs with \textcolor{black}{hyperbolic and radial nature in the displacement vector field at $\gamma=0.099$ are shown in two dimensional projections ($xy$, $yz$ and $xz$ planes) in magenta (hyperbolic) and green (radial), respectively. The yellow regions are the plastic soft spots, identified using the softness parameter, Eq.~\eqref{softdef}. {\bf d} The 3D configuration with defects locations and soft spots at $\gamma=0.099$ are shown. {\bf e} The average spatial correlation between monopoles (in the displacement field) and soft spots and anti-monopoles (in the displacement field) and soft spots. {\bf f} The average spatial correlation between hyperbolic defects (in the displacement field) and soft spots and radial defects (in the displacement field) and soft spots are shown in magenta and green respectively. The average is calculated over several $\gamma$ values corresponding to sudden stress drops, and error bars are included.}}
    \label{fig5}
\end{figure*}

We then move on to perform the same analysis using method (II), based on the non-affine displacement magnitude. \textcolor{black}{We present a snapshot at $\gamma = 0.099$, highlighting the locations of soft spots and HTDs with hyperbolic and radial geometries. In Figure~\ref{fig5}a-c, we show 2D projections in the $xy$, $yz$, and $xz$ planes, respectively, along with a full 3D view in Figure~\ref{fig5}d. The strong spatial correlation between the soft spots and HTDs is evident.}

We \textcolor{black}{compute} the spatial correlation function between monopoles in the displacement field and soft spots (\color{red}\textbf{red} \color{black} data) and between anti-monopoles in the displacement field and soft spots (\color{blue}\textbf{blue} \color{black} data). \textcolor{black}{In Figure~\ref{fig5}e, we present the averaged correlation functions between soft spots (method II) and HTDs with topological charges $Q = +1$ and $Q = -1$, averaged across 10 different plastic events, as shown in Figure~\ref{figSI1}c. Additionally, in Figure~\ref{fig5}f, we show the averaged correlation functions between soft spots (method II) and HTDs classified by hyperbolic and radial nature.}

We observe a very strong spatial correlation between the HTDs in the displacement field and the soft spots, with a pronounced peak appearing at $r \approx 1$. Interestingly, the distance at which this correlation emerges is compatible with the low-frequency results presented in Figure~\ref{fig2}, where the HTDs have been identified in the eigenvector field, and the results in Figure~\ref{fig3} as well. This confirms our initial speculation that, around plastic events, the topology of the low-frequency eigenvectors, and even its spatial characteristics, are compatible, and in fact very similar, to those observed in the displacement vector. 

Finally, we notice that there is no apparent difference between negative ($Q=-1$) and positive ($Q=+1$) HTDs, or at least in their correlation with plasticity. This is independent of the methods employed to define the HTDs and also the soft spots, and therefore rather universal. Interestingly, this represents a dissimilarity with respect to the analysis using the eigenvector field in 2D (or in the 2D-slicing method for 3D systems), where negative defects seem to have a privileged role \cite{wu2023,desmarchelier2024topological,bera2024soft}. \textcolor{black}{When we classify the HTDs based on their geometric nature as hyperbolic or radial, we observe stronger correlations for hyperbolic defects. This trend holds consistently across all cases studied in this work.}

In addition, we notice that the results shown in Figure~\ref{fig4}, using the non-affine displacement field, display a rather long-range structure in which spatial correlations survive up to large distances, $r \approx 5$. This very long correlation length might depend on several parameters and deserves further investigation. Moreover, the spatial structure of the HTDs and the spatial decay of their correlation with soft spots might provide useful insights to achieve a microscopic definition of the elastic screening length proposed in other mesoscopic approaches to amorphous solids \cite{PhysRevE.104.024904}.

\section*{Discussion} 
In this work, we proposed to use point-like hedgehog topological defects (HDTs) to characterize the plasticity of 3D amorphous solids and to locate their soft spots using well-defined mathematical concepts. First, borrowing from the physics of liquid crystals \cite{Kleman,kleman12006,holbl2023} and Heisenberg magnets \cite{fumeron2023introduction,lau1989,holm1994,berg1981}, we have outlined a concrete and practical procedure to identify HDTs in 3D glasses -- a method that is suitable for experimental discrete data sets as well.

We have then applied this proposed algorithm to both the eigenvector field and the displacement vector of a simulated 3D polymer glass model and verified the feasibility of this method. For the HTDs identified in the eigenvector field, we have proved that the number of HTDs as a function of frequency follows the same trend as in previous 2D studies, and directly correlates with the vibrational density of states at low frequency. We have also shown a strong tendency in forming oppositely charged monopole/anti-monopole pairs of topological defects and even clusters with larger number of defects. Moreover, we have revealed a very strong spatial correlation between the hedgehog topological defects and the plastic soft spots at short distances.

Finally, we considered HTDs in the 3D displacement vector field for a quasi-static shear deformation and for different large plastic events, including the yielding point. In this analysis, we have found that HTDs correlate very strongly with the extended plastic soft spots from which plastic instabilities originate. \textcolor{black}{These soft spots are identified using two different methods: one based on the softness parameter \cite{C4SM01438C} and the other on the amplitude of the non-affine displacement field. This result provides strong evidence that the location of HTDs, along with their yet undisclosed dynamics, may play a pivotal role in achieving a microscopic understanding of plasticity and yielding in 3D glasses.}

Interestingly, the analysis of topological defects in 3D amorphous solids revealed that the correlation with the plastic soft spots is independent of the charge of the defect, unlike previous findings in 2D systems \cite{wu2023,desmarchelier2024topological,bera2024soft} and in the 2D-slicing of 3D systems \cite{bera2024soft}. \color{black} This results can be directly understood from the arbitrariness in the sign of the topological charge $Q$.

On the other hand, we find that in 3D, the geometrical properties of the point defects play a fundamental role in identifying the microscopic structures responsible for plasticity. In particular, we observe that hedgehog topological defects with hyperbolic nature are ultimately the responsible for plastic deformations in 3D amorphous solids. This aligns well with previous findings in 2D \cite{wu2023}, where the corresponding carriers are the ones with negative winding number (anti-vortices). The connection between these two objects is that they both share a saddle-like geometrical structure that is more prone to plastic instabilities and hence strongly correlates with the plastic spots. \color{black}

Before concluding, we also notice that the correlation between plasticity and topological defects appears stronger when the latter are defined using the displacement field, rather than the low-frequency eigenvector fields. This is promising since, apart from few exceptions (\textit{e.g.}, \cite{vaibhav2024experimental}), the eigenvector field is not a quantity directly accessible via experiments (with the exception of colloidal systems \cite{vaibhav2024experimental}), whereas the displacement vector is. \textcolor{black}{Although there may be a connection between the structural properties of the material and these HTDs, none of the recent studies have explicitly reported it. See \cite{liu2024measurable} for a recent attempt in this direction based on inversion-symmetry and \cite{huang2024spotting} for a study in defective crystals.}
 
On a general ground, our work proves that the idea of using geometry and topology to describe, predict and rationalize plastic properties of amorphous solids can be successfully extended to three dimensional systems, such as real materials. Aside from the theoretical part of this development, this result provides a formidable occasion to test these ideas on realistic experimental amorphous solids for which the method we proposed can be directly applied. For example, three dimensional sheared granular \cite{Cao2018} and colloidal \cite{chikkadi} systems constitute possible candidate platforms to test these ideas.
\section*{Methods}
\subsection*{Model}
~~In this study, we implement the Kremer-Grest model \cite{Kremer} to simulate a coarse-grained polymer system. This system comprises linear polymer chains, each consisting of 50 monomers, with alternating monomer masses of $m_1 = 1$ and $m_2 = 3$ \cite{PhysRevB.102.024108}. The total number of monomers in our system was $N = 10{,}000$.

\begin{figure}[ht]
    \centering
\includegraphics[width=0.85\linewidth]{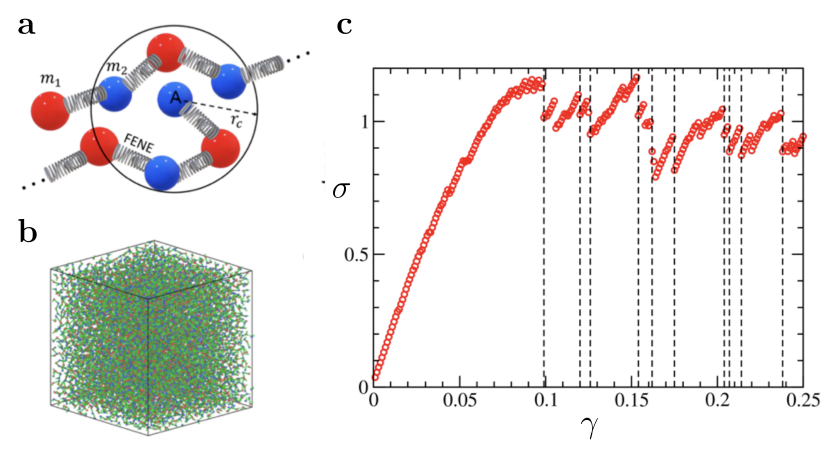}
    \caption{{\bf Simulation model and stress versus strain curve.} \textbf{a} Schematic representation of the polymer chains with alternating masses $m_1$ and $m_2$. The FENE bonds and the LJ interactions range $r_c$ from a monomer (A) are shown. \textbf{b} A typical simulation configuration is presented where monomer breads with masses $m_1$ and $m_2$ are indicated in red and blue, respectively and FENE bonds by green colors. \textbf{c} Stress ($\sigma$) versus strain ($\gamma$) curve. \textcolor{black}{Several stress drops, indicated as plastic events, are marked by black dashed lines. The yielding point ($\gamma=0.099$) and other plastic events at $\gamma=0.012$, $0.126$, $0.154$, $0.162$, $0.175$, $0.204$, $0.207$, $0.214$, and $0.238$ are analyzed separately, and the averaged correlation results are presented in Figure~\ref{fig4}a-b and Figure~\ref{fig5}e-f in the main text.}}
    \label{figSI1}
\end{figure}

The interactions between monomers in our model are governed by the truncated and shifted Lennard-Jones (LJ) potential:
\begin{equation}\label{full_lj}
u_{\text{LJ}}(r)=
\begin{cases}
V(r)-V(r_c), & \text{for}~r<r_c=2.5\sigma,\\
0, & \text{otherwise.}
\end{cases}
\end{equation}
Here, $V(r)$ represents the standard Lennard-Jones potential:
\begin{equation}\label{lj}
V(r) = 4{\varepsilon} \left [ \left (\frac{\sigma}{r} \right )^{12} - \left (\frac{\sigma}{r} \right )^6 \right ],
\end{equation}
where $\varepsilon$ is the interaction strength, and $\sigma$ is the monomer diameter. Here $r_c = 2.5\sigma$ is the interaction distance.

In addition to LJ interactions, covalent bonds along the polymer chains are modeled using the finite extensible nonlinear elastic (FENE) potential:
\begin{equation}\label{fene}
u_{\text{FENE}}(r) = -\frac{Kr_0^2}{2} \ln \left [ 1 - \left (\frac{r}{r_0} \right )^2\right ].
\end{equation}
We set $K = 30$ and $r_0 = 1.5\sigma$ in our simulations. In Figure~\ref{figSI1}a we schematically show these interactions. The polymer chains were simulated within a cubic box with periodic boundary conditions applied in all three dimensions. The system was equilibrated using the LAMMPS simulation package \cite{lammps}. We show a typical configuration in Figure~\ref{figSI1}b. Following equilibration, we employed an athermal quasi-static (AQS) deformation protocol. This involved quenching the glass sample to absolute zero temperature and subsequently applying quasi-static shear with a strain increment of $\delta \gamma_{xz} = 0.001$. The units of mass, length, and time of our simulation are taken as $m_1$, $\sigma$, and $\tau = \sqrt{m_1\sigma^2/{\varepsilon}}$, respectively. We set $m_1$, $\sigma$, and $\varepsilon$ to unity. Throughout our simulations, polymer chains were considered fully flexible, with no angle-bending terms included in the potential. The stress-strain curve in the shear plane is presented in Figure~\ref{figSI1}c. 

\subsection*{Numerical methods and data analysis}
\begin{figure}[ht]
    \centering
    \includegraphics[width=0.5\linewidth]{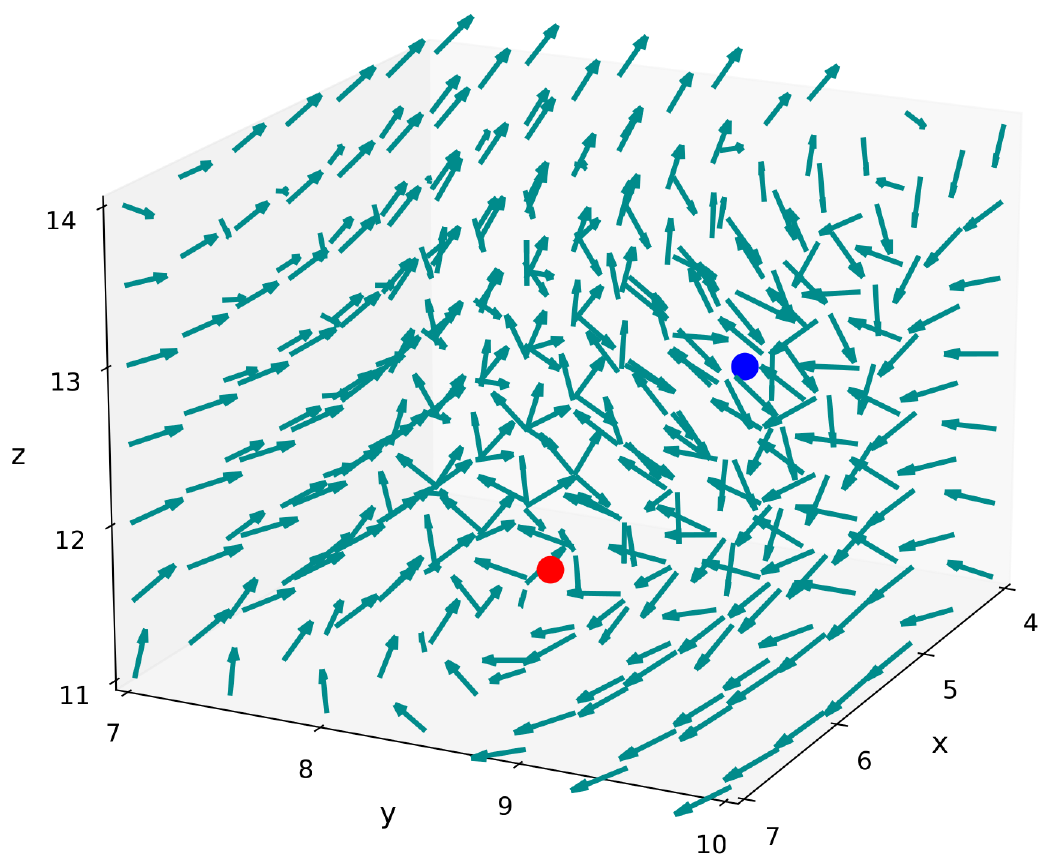}
     \caption{Normalized eigenvector field for a portion of the simulation box exhibiting a monopole defect (\color{red}\textbf{red}\color{black}) and an anti-monopole one (\color{blue}\textbf{blue}\color{black}). These data correspond to $\omega=3$.}
    \label{figSI2}
\end{figure}

We diagonalize the dynamical matrix as described in Eq. \ref{eq_hessian} to obtain the eigenvalues and eigenvectors. For each eigenvector, represented as a $3N \times 1$ column matrix corresponding to an eigenfrequency $\omega_k$ ($k = 1, 2, \ldots, 3N$), we assign the eigenvector field ${\mathbf{e}}_i = (e^x_i, e^y_i, e^z_i)$ to each particle position $\vec{r}_i$ ($i = 1, 2, \ldots, N$). This eigenvector field is interpolated onto a $40 \times 40 \times 40$ simple cubic lattice with grid spacing of $\sigma/2$ in all directions, superimposed on the 3D simulation box. For each eigenfrequency, the eigenvector field at each lattice site $\vec{r}=(l,m,n)$ is then obtained as \cite{wu2023}
\begin{equation}
   e^{\mu}(\vec{r}) = \frac{\sum_{d_i < R_{\text{cut}}} w(\vec{r} - \vec{r}_i)e^{\mu}_i}{\sum_{d_i < R_{\text{cut}}} w(\vec{r} - \vec{r}_i)}, \quad \mu = x, y, z,
\end{equation}
where $\vec{r}_i$ is the position of particle $i$, and $w(\vec{r} - \vec{r}_i)$ is a Gaussian weight function given by $w(\vec{r} - \vec{r}_i) = \exp(-|\vec{r} - \vec{r}_i|^2 / r_c^2)$ with $r_c = 1$. The distance $d_i = |\vec{r} - \vec{r}_i|$ is the separation between the lattice site and particle $i$, and $R_{\text{cut}} = 4\sigma$ is the cutoff distance used for interpolation. We employ this cutoff radius, $R_{\text{cut}}$, to accelerate the simulation. We tested the robustness of the interpolated field by considering larger cutoff distances and the full long-range limit. The eigenvector field at each lattice grid point is then normalized and used to identify the hedgehog topological defects (HTDs). \textcolor{black}{In Figure~\ref{figSI2}, we show a snapshot of a 3D eigenvector field presenting two clear topological defects with opposite charge at eigenfrequency $\omega=3$.}

In order to compute the topological charge density, we divide the entire 3D simulation box into cubic cells of side $\sigma/2$. Within each such cell we count number of HTDs $N^+$ and $N^-$ with $Q=+1$ and $Q=-1$, respectively, for $\omega_k < 4$. The topological charge density inside each cubic cell is defined as
\begin{equation}
    \bar{C}=\frac{\sum_{\omega_{k}} (N^+-N^-)/{\omega}_k^2}{\sum_{\omega_k} 1/{\omega}_k^2},
\end{equation}
where the weight factor is motivated by the quadratic dependence of the number of HTDs at low frequency.

We derive the displacement field as \cite{baggioli2021,bera2024soft} $\vec{u}_i={\vec{r_i}}(\gamma)-{\vec{r}_i}(\gamma-\delta\gamma)$, where $\vec{r}_i(\gamma)$ and $\vec{r}_i(\gamma-\delta\gamma)$ represents the locations of the particle $i$ at strains $\gamma$ and $\gamma-\delta\gamma$, respectively. The affine component for the simple shear scenario is calculated as $\vec{u}_{A} = z\delta\gamma \hat{x}$. Finally we obtain the non-affine displacement field by subtracting the affine component from the total displacement, $\vec{u}_{\text{NA}}=\vec{u}_i-\vec{u}_{A}$ \cite{baggioli2021,bera2024soft}. To obtain HTDs within the non-affine displacement field, we interpolated the discrete non-affine displacement field $\vec{u}_{\text{NA}}$ onto a simple cubic lattice grid with a side length $\sigma/2$ using the standard Python package \textit{griddata} with linear interpolation.
\section*{Data availability}
The datasets generated and analysed during the current study are available upon reasonable request by contacting the corresponding authors. 

\section*{Code availability}
The codes that support the findings of this study is available upon reasonable request by contacting the corresponding authors. 

\section*{Acknowledgements}
 M.B. thanks W.~Kob, Z.~Wu, J.~Zhang, Y.~Wang, C.~Jiang, Z.~Zheng, H.~Tong for useful discussions. We would like to thank A.~Liu and T.~Petersen for ongoing collaborations and endless discussions about this topic as well. M.B. acknowledges the support of the Shanghai Municipal Science and Technology Major Project (Grant No.2019SHZDZX01) and the sponsorship from the Yangyang Development Fund. A.Z. gratefully acknowledges funding from the European Union through Horizon Europe ERC Grant number: 101043968 ``Multimech''. A.Z. gratefully acknowledges the Nieders{\"a}chsische Akademie der Wissenschaften zu G{\"o}ttingen in the frame of the Gauss Professorship program. A.Z. and A.B. gratefully acknowledge funding from US Army Research Office through Contract No. W911NF-22-2-0256.

\section*{Author contributions}
A.~B. performed the numerical computations and data analysis. M.~B. and A.~B. wrote the manuscript with the help of A.~Z. All Authors contributed to the physical interpretation of the results.

\section*{Competing interests}
The authors declare no competing interests.

\section*{Supplementary Material}
\subsection*{Number of HTDs and vibrational density of states}
\begin{figure}[h]
    \centering
\includegraphics[width=\linewidth]{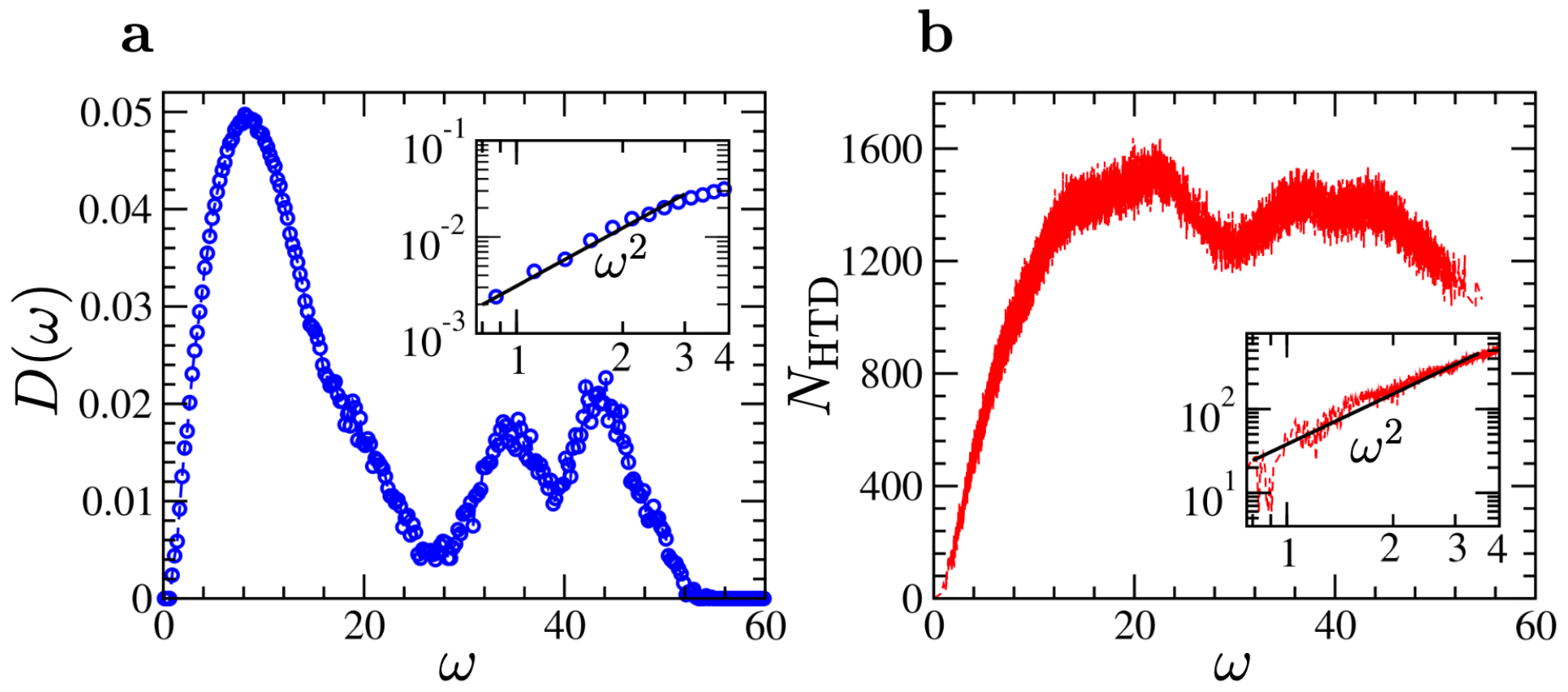}
    \caption{{\bf The vDOS and the number of HTDs versus eigenfrequencies.} \textbf{a} The vibrational density of states $D(\omega)$ of the polymer glass model at $T=0$ extracted from the direct diagonalization of the Hessian matrix. The inset highlights the Debye scaling of the vDOS at low frequency. \textbf{b} The number of hedgehog topological defects in the eigenvector field $N_{\text{HTD}}$ as a function of frequency $\omega$. The inset zooms on the low frequency region and highlights the $\omega^2$ scaling.}
    \label{figSI3}
\end{figure}

In Figure~\ref{figSI3}a, we present the vibrational density of states (vDOS) of the polymer glass model at $T=0$. As highlighted in the inset, at low frequency the vDOS exhibits a distinct Debye scaling, $D(\omega) \propto \omega^2$. In panel b of the same figure, we present the number of hedgehog topological defects in the eigenvector field $N_{\text{HTD}}$ as a function of frequency $\omega$. At low frequency, the number of defects also scales as $\omega^2$, implying a correlation with the vDOS.

\begin{figure}[h]
    \centering
\includegraphics[width=0.7\linewidth]{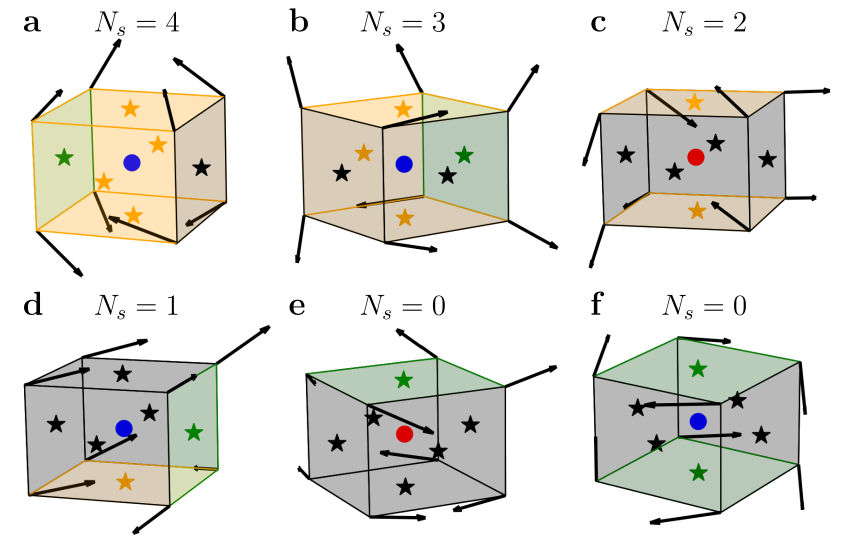}
    \caption{\textcolor{black}{{\bf 3D defect structures for different $N_s$.} \textbf{a–f}, Snapshots of the 3D vector field on cubic cells. Defects with topological charge $Q=+1$ ($Q=-1$) at the center are shown in red (blue). The winding numbers $q_{2D}$ of the projected vector field on the cube’s faces are marked with stars: $q_{2D}=+1$ in green, $q_{2D}=-1$ in orange, and $q_{2D}=0$ in black. Each face is colored according to the corresponding 2D point defect. The panels illustrate various configurations with different values of $N_s$, the number of faces exhibiting $q_{2D}=-1$.}}
    \label{figSI4}
\end{figure}

\begin{figure}[h]
    \centering
\includegraphics[width=\linewidth]{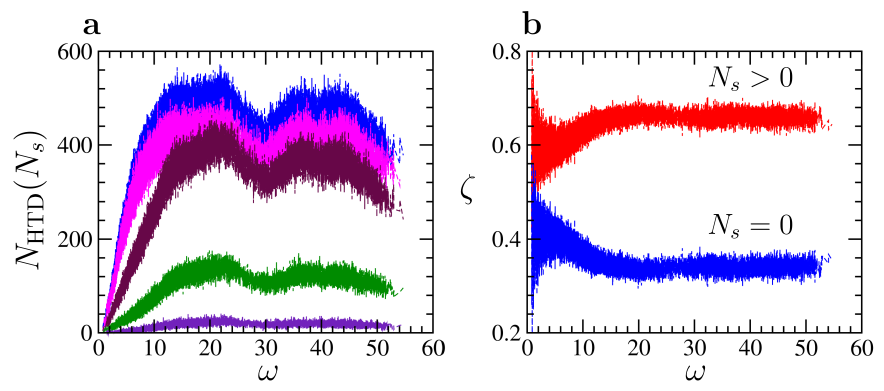}
    \caption{\textcolor{black}{{\bf Behavior of HTDs with different saddle-surfaces.} \textbf{a} The number of hedgehog topological defects, $N_{\text{HTD}}(N_s)$, as a function of eigenfrequency $\omega$,  for different values of $N_s$ (the number of saddle-surfaces). Data are shown in distinct colors for $N_s=0$ (blue), $N_s=1$ (magenta), $N_s=2$ (maroon), $N_s=3$ (green) and $N_s=4$ (indigo). \textbf{b} The fraction of HTDs, $\zeta=N_{\text{HTD}}(N_s)/N_{\text{HTD}}$ plotted against $\omega$, for radial ($N_s=0$) and hyperbolic ($N_s>0$) defects.}}
    \label{figSI5}
\end{figure}

\subsection*{\textcolor{black}{Hyperbolic and radial HTDs with varying saddle-surfaces}}
\textcolor{black}{To classify hyperbolic and radial defects, we project the three-dimensional vector field onto multiple two-dimensional planes and compute the winding number, $q_{2D}$, on the six faces of a cubic cell. The winding number takes discrete values of $+1$, $0$, and $-1$, corresponding to different topological structures. In Figure~\ref{figSI4}, we present several 3D defect structures along with their 2D projections on the faces of cubic cells. Each face of the cube is color-coded based on its $q_{2D}$ value: green for $q_{2D}=+1$, black for $q_{2D}=0$, and orange for $q_{2D}=-1$. We define $N_s$ as the number of cube faces having $q_{2D}=-1$, representing the number of saddle-like surfaces. Here, we show several vector field configurations on cubic cells, highlighting different values of $N_s$ and their corresponding defect structures.}

\subsection*{\textcolor{black}{Behavior of HTDs with different $N_s$ as a function of eigenfrequency}}
\textcolor{black}{In Figure~\ref{figSI5}a, we analyze the frequency dependence of the number of HTDs, $N_{\text{HTD}}(N_s)$, for different values of $N_s$. Our results indicate that $N_{\text{HTD}}(N_s)$ decreases with increasing $N_s$, suggesting that defects with multiple saddle-like projections occur less frequently. We compute the ratio $\zeta$, which represents the fraction of HTDs with a given $N_s$ relative to the total number of HTDs. In Figure~\ref{figSI5}b, we present this ratio separately for hyperbolic defects ($N_s>0$) and radial defects ($N_s=0$). At low frequencies, the difference in the number of hyperbolic and radial defects is small, while at higher frequencies, hyperbolic defects become more prevalent than radial defects.}

\subsection*{\textcolor{black}{Correlation between soft Spots and HTDs with hyperbolic and radial nature}}  
\begin{figure}
    \centering
  \includegraphics[width=0.85\linewidth]{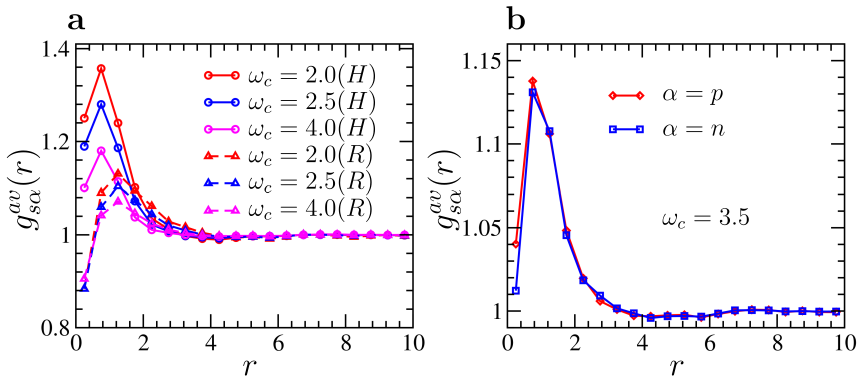}
    \caption{\textcolor{black}{{\bf Average correlations between soft spots and HTDs.} \textbf{a} Average spatial correlation between soft spots (s), identified using the softness parameter (Eq.~\eqref{softdef}), and hyperbolic (H) and radial (R) defects for different frequency cutoffs: $\omega_c = 2.0$, $2.5$, and $4.0$. The correlation with hyperbolic defects remains consistently stronger than with radial defects across all $\omega_c$.  
    \textbf{b}, Average correlation between soft spots (s) and hedgehog topological defects (HTDs) with positive (p) and negative (n) topological charges, showing no significant difference between the two cases.}}

    \label{figSI6}
\end{figure}

\textcolor{black}{We analyze the average correlation between soft spots and hyperbolic (H) defects, as well as soft spots and radial (R) defects, for different frequency cutoffs ($\omega_c$), as shown in Figure~\ref{figSI6}a. Our results indicate that the correlation with hyperbolic defects is consistently stronger than with radial defects across all $\omega_c$. Furthermore, we observe a systematic decrease in the peak height with increasing $\omega_c$, suggesting that the correlations are strongest at lower frequencies. In Figure~\ref{figSI6}b, we present the average correlation between soft spots and positive (p) HTDs, as well as soft spots and negative (n) HTDs. No significant differences are observed between these cases, consistent with the discussion in the main text.}

\subsection*{Extended data and correlation between soft spots identified with different methods}
\begin{figure}
    \centering
    \includegraphics[width=0.7\linewidth]{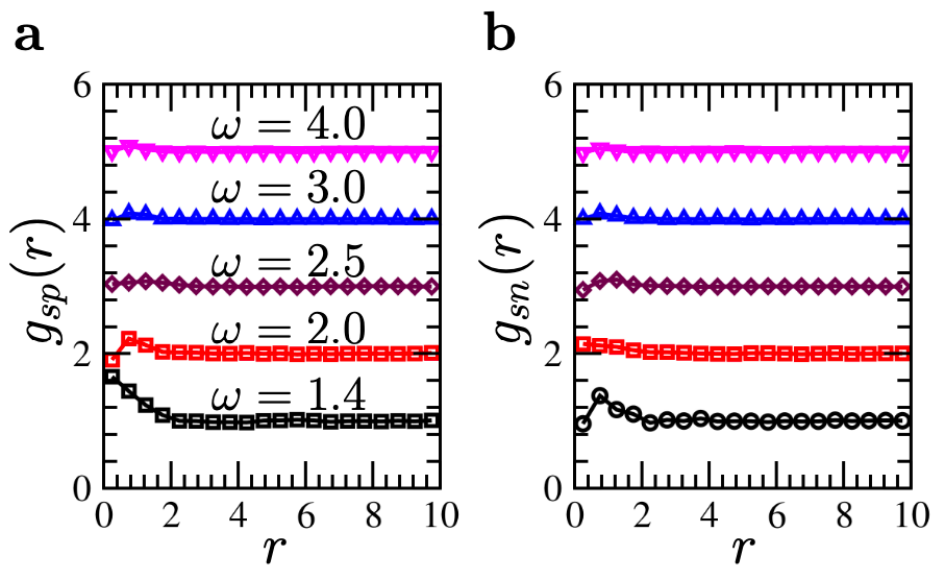}
    \caption{\textcolor{black}{{\bf  Extended data on soft spots and HTDs with positive and negative charge.}} The spatial correlation between soft spots (s) identified using the softness parameter, Eq.~\eqref{softdef}, and negative (n) and positive (p) HTDs for several higher frequencies. See Fig.~\ref{fig2} in the main text for low-frequency data.}
    \label{figSI7}
\end{figure}

In Figure~\ref{figSI7}, we provide extended data for the spatial correlation between between soft spots (s) identified using the softness parameter, Eq.~\eqref{softdef}, and negative (n) and positive (p) HTDs. Low-frequency data are provided in Fig.~\ref{fig2} in the main text. From Figure~\ref{figSI7}, the vanishing of the correlation at high frequency ($\omega \gtrapprox 2$) is evident.

\begin{figure}
    \centering
  \includegraphics[width=0.5\linewidth]{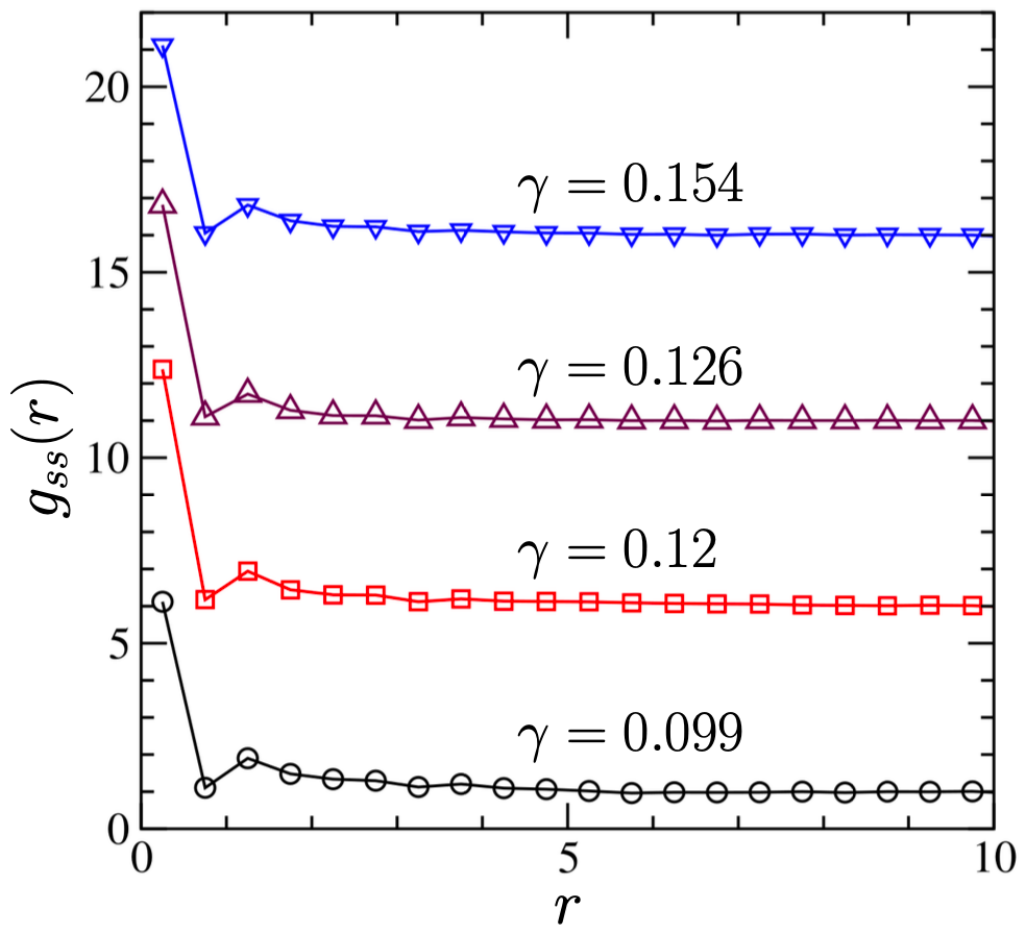}
    \caption{\textcolor{black}{{\bf Correlations between soft spots using method I and method II.}} Spatial correlation between soft spots identified using the softness parameter, Eq.~\eqref{softdef}, and the non-affine displacement field. Different lines correspond to different values of strain, located around major plastic events along the stress-strain curve, Figure~\ref{figSI1}c.}
    \label{figSI8}
\end{figure}

Finally, for completeness, in Figure~\ref{figSI8} we show the spatial correlation between the plastic soft spots identified with method I (softness parameter, Eq.~\eqref{softdef}) and method II (non-affine displacement field). From there, we observe that the two definitions provide a strong correlation at short distance that, nevertheless, does not survive for large length scales.

\end{document}